# Empirical Model for Discharge Current and Momentum Injection in Dielectric Barrier Discharge

Anthony Tang[1,*], Ravi Sankar Vaddi[1,*], Alexander Mamishev[2], Igor V Novosselov[1,3,†]

[1]Department of Mechanical Engineering, University of Washington, Seattle, U.S.A. 98195

[2]Electrical and Computer Science Department, University of Washington, Seattle, U.S.A. 98195

[3]Institute for Nano-Engineered Systems, University of Washington, Seattle, U.S.A. 98195

## ABSTRACT

Dielectric barrier discharge (DBD) plasma actuators with an asymmetric, straight edge electrode configuration generate a wall-bounded jet without moving parts. Mechanistic description of the interaction between the Coulombic forces and fluid motion as a function of DBD parameters remains unclear. This paper presents an experimental investigation of DBD actuator, including electrical current associated with microdischarges, plasma volume, and the wall jet momentum over a range of AC frequencies (0.5 – 2 kHz) and peak-to-peak voltages up to 19.5 kV. Discharge current is measured with a high temporal resolution, and plasma volume is characterized optically, and the momentum induced by the DBD wall jet is computed based on the axial velocities measured downstream of the actuator using a custom-built pitot tube. Discharge current analysis demonstrated asymmetry between the positive and negative semi-cycle; both currents yielded a power-law relationship with empirical fitting coefficients. Plasma length varies linearly and volume quadratically with voltage. Although plasma length reached an asymptotic value at a higher frequency, the plasma volume grows due to the increasing height of the ionization region. In a simple 2D configuration, the DBD wall jet momentum shows near-linear dependency with discharge current in the range of voltages and frequencies considered in this work. The presented empirical model characterizes the DBD wall jet momentum and the discharge current based only on the AC inputs. With the estimation of plasma volume, the model can be applied for determining more realistic boundary conditions in numerical simulations.

Keywords: Empirical Model, Dielectric Barrier Discharge, Wall Jet, Plasma Volume, Discharge Current, Momentum Injection

## 1. INTRODUCTION

Over the past decade, there has been a great interest in using non-thermal plasma actuators for active flow control of aerodynamic surfaces [1-7]. Plasma actuators have the potential to control a fluid system while staying silent, instantaneous, and compact [8-10]. Generated through corona discharge or dielectric barrier discharge (DBD), the ions are generated when an applied voltage induces an applied electric field that exceeds the dielectric strength of air or any working fluid. The interaction between free ions, accelerated by E-field, working fluid, and surfaces can be utilized in applications such as aerodynamic drag reduction [11-13] and electric propulsion [14-17]. Still, despite their lower electromechanical efficiencies, than corona driven flow, DBD actuators are more effective at providing a consistent electro-hydrodynamic (EHD) force [4, 9]. The current DBD applications are primarily limited to flow control at low-speed conditions due to their relatively lower EHD forces [14, 18, 19]. There are several fundamental aspects of DBD that are not well-understood such as the kinetics of ion recombination, ionization of different species, and coupling between discharge and fluid flow. Thus, many studies have explored these multiphysics phenomena to optimize the electrical and mechanical effects of the DBD via analytical and numerical modeling [20-23]. The previous work in modeling corona EHD can be relevant [24, 25]. However, the transient nature of AC discharge significantly complicates modeling as it involves several time scales, such as electrical frequency, ionization kinetics, electron transport in the dielectric medium, fluid advection, etc. An empirical model that can describe the relationship between the applied AC (voltage and frequency) to discharge current and DBD wall jet properties could be very beneficial from a practical perspective.

[*] These authors contributed equally to this work

[†] ivn@uw.edu

This study explores the relationship between the AC inputs and electrical and fluid mechanical properties of a straight-edge DBD plasma actuator. The actuator comprises two electrodes separated by a thin dielectric, as shown in Figure. 1. Known as the active electrode, an air-exposed high voltage copper film rests on the dielectric surface while the other electrode is encapsulated in the dielectric and grounded. When a high voltage is applied, the electric field is strongest in the region between the electrodes; the plasma is generated at the active electrode's edge [26-28]. The downstream length (or simply length) of the electrodes is usually a few millimeters, and different studies have explored the effects of gaps between the electrodes [5, 29, 30]. The effect of the thickness of the electrodes and the dielectric media play a significant impact on the actuator's performance [28, 31-34]. A single straight-edged DBD actuator is assumed to produce a two-dimensional flow field due to the spanwise uniform electric field. Other actuator designs have been considered, including serrated electrodes that produce a three-dimensional flow field [35-37]. The spanwise length of the electrodes serves as a nominal reference length in the analysis [5].

Traditional metrics to characterize plasma actuators' performance include current, velocity, and one-dimensional plasma length measurements. A current measurement through a non-intrusive current coil can be viewed as a superposition of low-frequency capacitive current, discharge current, and noise. The capacitive current is often filtered out or ignored because it corresponds to the transiently stored energy in the dielectric or air and not energy transferred to fluid motion [5]. On the other hand, the discharge current indicates the amount of charged species that can participate in the energy transfer to fluid motion. The discharge current comprises numerous peaks in the positive-going cycle due to streamer propagation with the addition of glow discharge during the negative-going cycle [38]. Further time-resolved measurement have shown that both semi-cycles contribute to EHD force; their relevant contributions is the topic of active scientific discussions [5, 27, 39]. Directly measuring discharge current for both semi-cycles can shed insight into their roles in the momentum transfer. Velocity measurements are often obtained by pitot tubes or particle imaging velocimetry (PIV); these measurements are used to characterize momentum transferred from charged species to neutral molecules. Ion generation is accompanied by electron emissions, which can be observed as a purple glow and characterized by ionization zone length and height. Considering 2D geometry, the volume of ionization can be calculated from length and height measurements. The size and the charge density in the ionization zone can be considered analogous to virtual origin parameters in the wall jet literature [40-42]. DBD wall similarity analysis was recently proposed [43]; however, additional experimental data is needed to perform robust non-dimensional analysis.

Despite the complex transient nature of DBD discharge, the plasma volume can serve as the bridge for understanding the coupling between the electrical and fluid mechanical characteristics. In developing an empirical model, one has to investigate the coupling between electrical and fluid mechanical properties of the actuator without modeling the complex chemistry and species transport [44]. The effects of electrical input by introducing the EHD body force term into Navier-Stokes equations. This term is expressed as

$$\vec{f}_{EHD} = \rho_c \vec{E}, \qquad (1)$$

where $\rho_c$ is the charge density and $\vec{E}$ is the electric field. The simplified models include the Orlov model [45], Shyy model [20], and Suzen & Huang model [46, 47]; these can significantly reduce the calculation times, especially for complex scenarios or 3D geometries, such as boundary-layer separation control, turbine blades, and channel flow [48-51]. Considering relative permittivity properties of the working fluid and actuator materials, the simplified DBD models define the charge density as a one-dimensional boundary condition with a half Gaussian distribution starting at the edge of the ground electrode closest to the active electrode [44]. The half Gaussian charge distribution has been experimentally and numerically investigated, and it is typically presented as a function of x-direction [50, 52-56]. The y-direction charge density distribution is indirectly considered through the Debye length [57, 58]. This assumption may not be appropriate for some condition, and thus, it require further investigation.

In this manuscript, we present an empirical model for an asymmetric low-profile DBD actuator based experimental data for discharge current and velocity measurements. Discharge current is reported for both positive and negative semi-cycles. The plasma volume and discharge current measurements are used to determine the charge density over range AC voltages and frequencies. Wall jet velocity measurements allow for analysis of the DBD momentum injection, which is then correlated to measured discharge current and inform a reduced-order empirical relationship. The electric to kinetic energy transfer efficiency is evaluated for the investigated conditions.

## 2. EXPERIMENTAL SETUP AND DIAGNOSTICS

### 2.1. DBD actuator

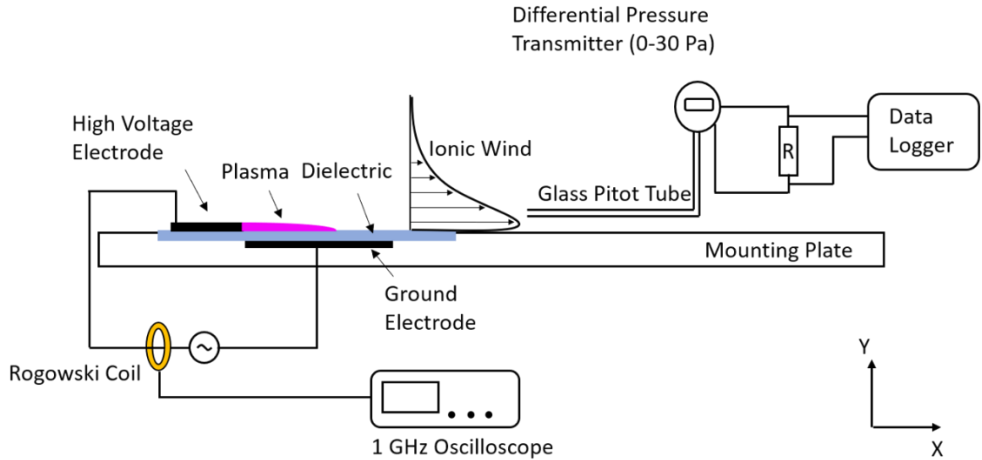

**Figure. 1.** Schematic of the dielectric barrier discharge (DBD) plasma actuator. The plasma actuator is mounted on an acrylic glass plate and the blue region is the dielectric layer separating the electrodes. The velocity is measured using a custom-built glass pitot tube and the electrical characteristics are measured using a Rogowski coil.

In the experimental setup, the electrodes are placed on opposite sides of the thin dielectric layer, in the asymmetric configuration, as shown in Figure. 1. Both electrodes have straight edges producing a uniform spanwise discharge, and thus, a 2D velocity profile. The dielectric material used in this study is Kapton (7700 VPM @ 25 °C). Each actuator has 1 layer of Kapton-FN (FEP layered Kapton) and 4 layers of Kapton-HN with a total thickness of ~330 μm (including the adhesive). The DBD actuator is installed on the 6″ by 8″ acrylic plastic sheet. The ground electrode (copper, 50 μm thick, 25 mm long, 110 mm wide) is flush-mounted on the acrylic plate. This encapsulation prevents plasma formation on the lower side of the dielectric material. The upper electrode (copper, 50 μm thick, 15 mm long, 110 mm wide) is glued onto the top of the Kapton dielectric layer. There is no overlap between the electrodes in the x-direction. The high-voltage (HV) electrode is exposed to atmospheric pressure air. The air-exposed HV electrode is connected to a Trek PM04014 power supply that provides up to 20 kV AC high voltage. The voltage and frequency were varied from 12 kV – 19.5 kV and 0.5 kHz – 2 kHz. This paper will use the coordinate system established in Figure. 1.

### 2.2. Plasma volume characterization

Optical measurements can be employed to reveal the light emitted during the plasma actuation, and it can be used to characterize the plasma area or projected plasma volume region. Previous studies have explored the one-dimensional plasma length and the plasma length's temporal evolution [5, 39, 54]. To measure the two-dimensional plasma discharge region, a Nikon D750 DSLR Camera with a Nikon AF-S NIKKOR 70 to 200 mm f/4G ED VR Zoom Lens is mounted horizontally 0.5 m from the side of the plasma actuator. The field of view has a resolution of 0.022 mm per pixel. To identify the discharge's volume, we first binarized the image with a 256-bit image histogram. Then, we use a 98% Otsu threshold to identify an effect plasma volume region. A 98% threshold matches previously determined one-dimensional plasma lengths, which have been experimentally and numerically validated [45, 57-59]. Typical lengths of plasma streamers range between 3 mm and 8 mm. However, further studies have shown that streamers can reach lengths up to 20 mm with thicker dielectrics [56, 58]. Figure. 2 below illustrates the plasma volume projection onto the x-y plane during the typical operation of the actuator. The threshold for this analysis is chosen to be 98% spectral intensity. Other values in the 95-99% range were evaluated, and the results are not affected significantly.

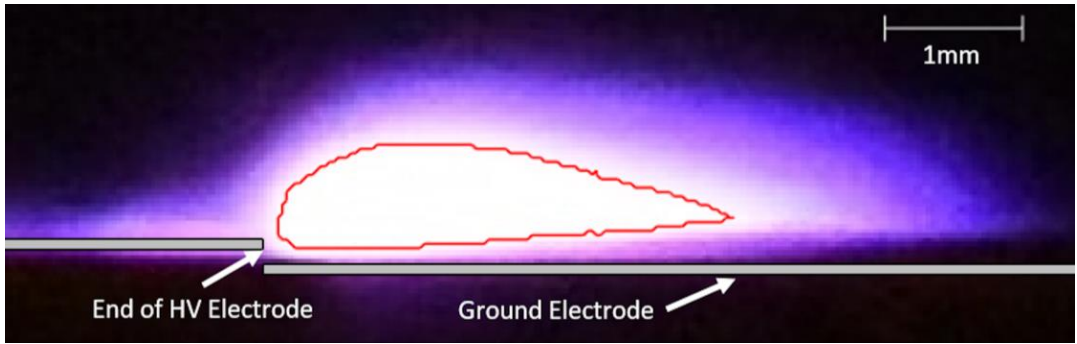

**Figure. 2.** Plasma discharge region of DBD actuators at 12 kV and 2 kHz. A 98% threshold is outlined in red to calculate the plasma length and volume.

The resulting identified plasma images are used to calculate the length, height, and projected volume of the discharge region. Note that the projected plasma volume has a sloped bottom boundary consistent in PIV measurements of EHD force distribution [26]. This is believed to be due to the surface charge effect on the dielectric surface.

### 2.3. Electrical measurements and signal processing

In a DBD, the electric current flowing into the circuit can be viewed as the superposition of a low-frequency capacitive current and a discharge current. The discharge current is associated with plasma microdischarges, and they appear as a series of fast current pulses [59], as shown in Figure. 3 (a). The current measurements are done using a 200 MHz bandwidth non-intrusive Pearson 2877 current monitor with a rise time of 2 ns. The current probe is placed around the wire driving the active electrode. The current monitor is connected to a Tektronix DPO 7054 oscilloscope that uses a bandwidth of 500 MHz to satisfy the Nyquist theorem by achieving a sampling rate of 1 GS/s. These conditions are essential for accurate capture of individual discharges that have been shown to occur on average over a 30 ns duration [60]. The high bandwidth and the sampling rate minimize the noise during the current measurements and can be used to compute the time-averaged electrical power [52]. A voltage from the power supply is also simultaneously displayed on the oscilloscope. The electrical power $W_{elec}$ consumed by the DBD is used to determine the actuator efficiency. It is derived by multiplying the voltage and current at each point in the time series and averaging over a entire period. The time-averaged electrical power consumed by the actuator can be computed as

$$W_{elec} = f_{AC} \int_{t^*=0}^{t^*=1} \varphi(t) \times i(t) \, dt, \qquad (2)$$

where $f_{AC}$ is the frequency of the applied voltage in Hz, and $\varphi(t)$ and $i(t)$ are respectively the voltage and current at each point in the period. The normalized $t^*$ represents a single period. We compute the averaged discharge current and resulting power from at least five separate periods to reduce the noise impact. Figure. 3 (a) below shows a typical DBD current measurement with a voltage curve. Notation for positive discharge (PD) and negative discharge (ND) indicate the semi-cycles when voltage is rising and voltage is dropping respectively.

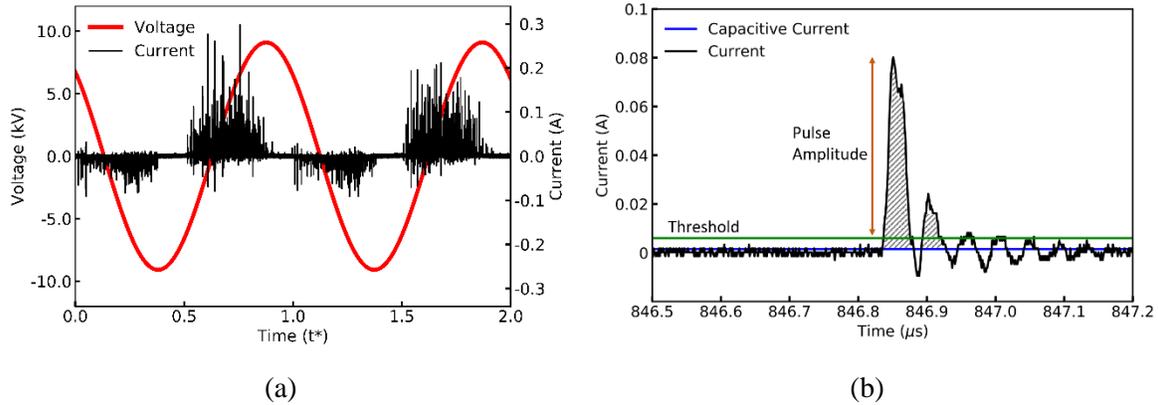

(a)  (b)

**Figure. 3.** (a) An example of DBD current with voltage signals at 18 kV (p-p) and 1 kHz applied frequency (b) Single microdischarge pulse shown with the underlying capacitive current (blue line) with two thresholds to characterize the microdischarge: (i) threshold of 6 mA (green line), (ii) the amplitude of the microdischarge (brown arrow) for distinguishing a microdischarge. The integration of the shaded area determines the charge associated with each discharge pulse.

To determine the current associated with the plasma microdischarges, we separate the capacitive current from the raw signal. Several have explored the capacitive current through analytical methods [58, 61]; others have removed the capacitive current through signal processing methods, including low-pass filters or Fast-Fourier Transform (FFT) [39, 59, 60]. In this study, we identify the capacitive current by considering the first 15 Fourier modes. The discharge current, the fast current events due to plasma microdischarges, appears as pulses emerging from a noisy baseline. Figure. 3 (b) shows a current peak along with the capacitive current and threshold to identify discharge current pulses. The capacitive current is used as a baseline to evaluate the intensity of the peaks. Two thresholds are introduced to characterize the microdischarge: (i) a threshold of 6 mA to evaluate the intensity of peaks, shown as a green line, and (ii) the length of the microdischarge of 10 mA represented by a brown line. This method allows to compute the charge delivered by the current pulse and discharge current per unit time.

The asymmetric electrode configuration has a similar structure to a corona discharge, though the transient behavior and the alternating field add significant complexity to the problem. For model development, an empirical relationship similar to Townsend's quadratic relationship can be used as a starting point to model discharge current [62, 63]. We can use the modified corona discharge energy consumption ($W$) equation to determine the discharge current of the actuator in Eq.(3)

$$W \sim f_{AC} C_0 (\varphi - \varphi_0)^2, \tag{3}$$

where $C_0$ is the capacitance of the electric circuit, $f_{AC}$ is the applied AC frequency, $\varphi$ and $\varphi_0$ are the applied and initiation voltage. We have assumed a power-law relation with frequency for the discharge current, as shown in Eq. (4)

$$I_{dis} \sim f_{AC}{}^\alpha C_0 (\varphi - \varphi_0)^2, \tag{4}$$

where $\alpha$ is the AC frequency effect on the discharge current. The capacitance of the DBD actuator with charges temporally on the surface is hard to determine, so Eq. (4) is rewritten as

$$I_{dis} = f_{AC}{}^\alpha K (\varphi - \varphi_0)^2 \tag{5}$$

where $K$ is the empirical constant similar to $C_1$ in Townsend current relationship $I_{dis} = C_1 \varphi (\varphi - \varphi_0)$. The empirical relation in Eq. (5), can be used to determine the expressions for total discharge current, PD and ND currents, and we can evaluate the microdischarge properties in both cycles. The value of $\varphi_0$ is determined from modified Peek's law [64] as shown in Eq. (6)

$$\varphi_0 = m_v g_v \left(\frac{t_e}{2}\right) \ln\left(\frac{2t_d + t_e/2}{t_e/2}\right),$$

$$m_v = 1, g_v = 31\left(1 + \frac{0.308}{\sqrt{t_e/2}}\right), \tag{6}$$

where $t_e$ and $t_d$ are the thickness of electrode and dielectric layer respectively. The value of $\varphi_0$ is calculated to be 4.75 kV and it is used for developing different power law relationships for discharge current, plasma volume and momentum induced by wall jet.

### 2.4. Wall jet characterization

The flow field induced by ions characterizes the fluid mechanical properties of the plasma actuators. To measure the time-averaged ionic wind velocity, we employ a custom-made glass pitot tube with a 0.4 mm inner diameter and 0.5 mm outside diameter. Compared to traditional stainless steel pitot tubes, the glass tube minimizes electrical interaction with the discharge. This method has been previously used to characterize plasma actuators' performance [5, 59, 65]. The pitot tube, mounted to an optical table and controlled in the x and y-axis, is connected to a Furness Controls FCO354 differential pressure transmitter (0 – 30 Pa). The pressure transmitter outputs a 4 – 20 mA current linear to its pressure range and is in series with a 1.5 kΩ resistor. The pressure within the pitot converges nearly instantly. The voltage across the resistor is recorded for a minimum of 30 seconds across a Hydra Data Logger II. With the time-averaged pressure ($P$), a time-averaged wind velocity ($v$) is calculated using Bernoulli's equation with a calibration correction factor ($C$) that is characteristic to custom pitot tube expressed as

$$\Delta P = C \rho v^2, \tag{7}$$

where $\rho$ is the fluid density. Using this configuration, the typical velocity measurements have a standard deviation of less than 0.02 m s$^{-1}$ over 30 s. In this experiment, only x-velocity measurements are taken at varying x positions (10 mm, 15 mm, 40 mm, and 75 mm) downstream on the active electrode edge. At each x position, the y-velocity profile is obtained from the surface to 6 mm above the plate at increments of 0.25 mm or 0.5 mm (at a higher location). Due to the geometry of the pitot tube, we cannot capture velocity at heights < 0.25 mm. As a result, we assume the velocity is linear between the no-slip condition at y = 0 and the velocity at y = 0.25 mm.

From a vertical velocity profile, the DBD actuator's total mass flow rate per meter spanwise, $Q$, can be computed by

$$Q = \rho \int_{y=0}^{y=\infty} U(y)dy, \tag{8}$$

where $U(y)$ is the measured velocity at varying heights at a constant x location. Similarly, the DBD actuator's total momentum per meter spanwise can be found by multiplying the mass flow rate at each vertical position with its respective velocity such that

$$M = \rho \int_{y=0}^{y=\infty} U^2(y)dy. \tag{9}$$

With no freestream flow and no external force, the momentum should theoretically be conserved in all velocity profiles. However, due to viscous forces and spatial charge effects, this is not the case in all profiles. The mechanical power of a given DBD plasma actuator corresponding to the kinetic energy ($W_{mech}$) in the actuator can be computed by

$$W_{mech} = \frac{1}{2}\rho L \int_{y=0}^{y=\infty} U^3(y)dy. \tag{10}$$

With both the electrical power as in Eq.(2) and mechanical power of a given actuator, the overall electromechanical efficiency of the plasma actuator can be calculated as

$$\eta = \frac{W_{mech}}{W_{elec}}. \tag{11}$$

## 3. RESULTS AND DISCUSSION

### 3.1. Effect of voltage and frequency on the plasma volume

In this section, the impacts of voltage and frequency on the effective plasma region are discussed. Figure. 4(a), presents the plasma length, measured from the edge of the plasma actuator to the furthest point downstream, as a function of voltage (12 kV – 19.5 kV) at varying frequencies (0.5 kHz – 2 kHz). The results support previously reported results, both experimental and numerical, that plasma length increases linearly with applied voltage [54, 58]. However, increases in frequency appear to asymptotically increase plasma lengths until 2 kHz. This asymptotic limit may correspond to the asymptotic limit between the max velocity and frequency at similar voltages where max velocities did not increase after frequencies larger than approximately 2 kHz [27]. However, Orlov reported the plasma length asymptotically approaches a limit at approximately 6 kHz while at 5 kV [58], thus there may be different asymptotic limits with varying dielectrics and electrode configuration. At our maximum conditions 19.5 kV and 2 kHz, the dielectric layer breakdown occurred after 60 min of operation, which was not often sufficient to take the entire set of the velocity data; thus, we did not explore higher frequency conditions. It was possible to operate the actuator at lower voltages and higher frequencies and then develop universal power-law equations. However, as we attempted to maximize the momentum injection, this exploration is out of the paper's scope, and further experimentation at a broader range of frequencies is necessary.

While the physical mechanisms such as ion recombination may play a role in limiting the plasma length, the thickness of the dielectric material has been previously noted to play a large role in plasma extension length. These results support previously reported lengths between 3 mm and 8 mm when the dielectric is less than 1 mm [5, 54, 56]. Thicker dielectrics have seen plasma lengths up to 20 mm, and thus the asymptotic nature of Figure. 4(a) may be due to the limiting length or thickness of the dielectric or the underlying ground electrode [5].

Figure. 4(b) presents the plasma volume, from the summation of pixels in the total effective plasma region, as a function of voltage (12 kV – 19.5 kV) at varying frequencies (0.5 kHz – 2 kHz).

Our results suggest that the volume varies quadratically with the applied voltage. Interestingly, the plasma volume continues to grow with frequency up to 2 kHz, whereas the plasma length approaches a limit at 2 kHz at these operating conditions due to continued growth in plasma height. The results suggest that the assumption of a space charge half-Gaussian distribution in the length and height hold for voltages and frequencies, which may not be the case at higher values. A change in plasma height would not be proportional to plasma length will likely lead to a change in the distribution of charges and a change in the resulting force and velocity profile.

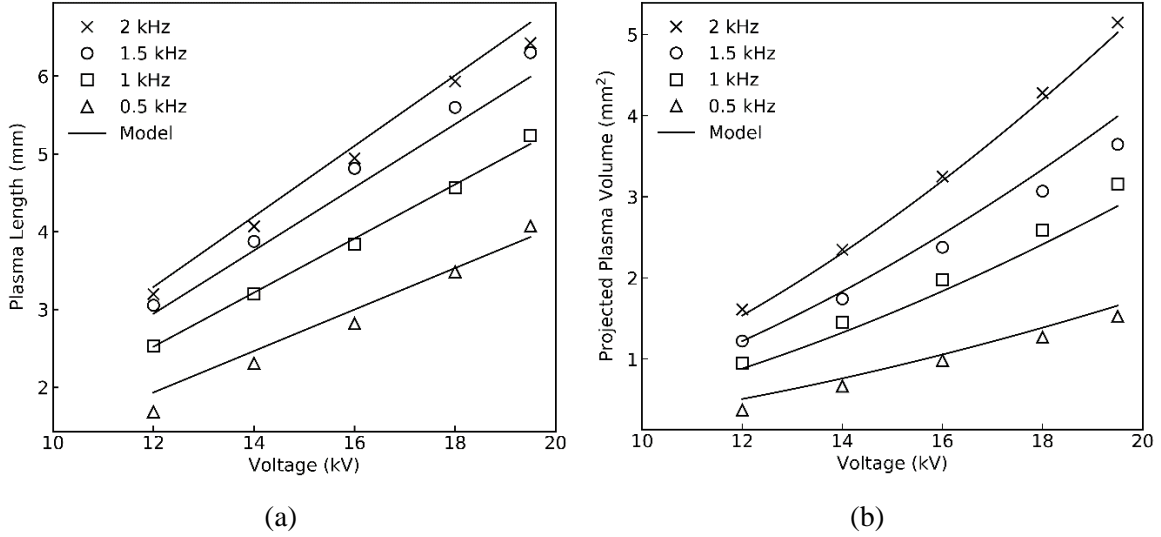

(a)  (b)

**Figure. 4.** Experimentally evaluated (a) plasma length and (b) projected plasma volume increase as a function voltage. Plasma length reaches an asymptotic value with frequency, but the plasma volume continues to increase.

The empical expressions for the plasma length and height can be determined through curve fitting. In this voltage range, plasma length ($L_P$) is found to vary linearly with voltage and by power-law with frequency as expressed in Eq. (12) and plasma volume ($V_P$) is found to exhibit a power-law relationship with both voltage and frequency as expressed in Eq. (13)

$$L_P = K_1 f_{AC}^{\alpha_1}(\varphi - \varphi_0), \tag{12}$$

$$V_P = K_2 f_{AC}^{\alpha}(\varphi - \varphi_0)^{\gamma}. \tag{13}$$

In numerical modeling for a source term, one needs to define the volume of charge injection. A gaussian ion-concentration distribution can then be applied to the empirical plasma length and height, and this relationship eliminates the dependency on a guessed Debye length. Subsequent coupling of Eq. (1) and the Navier-Stokes equation allows for the modeling of the EHD force term.

### 3.2. Discharge current characteristics

Temporally resolved current measurements allow characterizing the microdischarges for both the positive and negative cycles. During the microdischarge, the charged species are transported towards the electrodes, thus generating an electrical signal in the form of a pulse superimposed on a slow-moving capacitive current. The latter can be subtracted, and the current associated with microdischarges can be analysed. In the PD portion of the cycle the electrons move towards the exposed electrode, and in the ND the electrons move towards the ground. We calculate the charge transported in each microdischarge and the total transported charge by adding up the contribution of each current microdischarge; the total discharge current is shown in Figure. 3(b). The discharge current is normalized by a unit time. Previous work showed that a net charge of 40 nC is transported in a positive discharge cycle at 8.5 kV [39] and 10 nC is transported in a negative discharge cycle. As a comparison, we have observed a charge transport of 45 nC in PD and 15.3 nC in ND at 12 kV, 500 Hz. The discharge current is calculated for both PD and ND, and the net discharge current is calculated by adding both cycles. The experimental data is shown in Figure. 5.

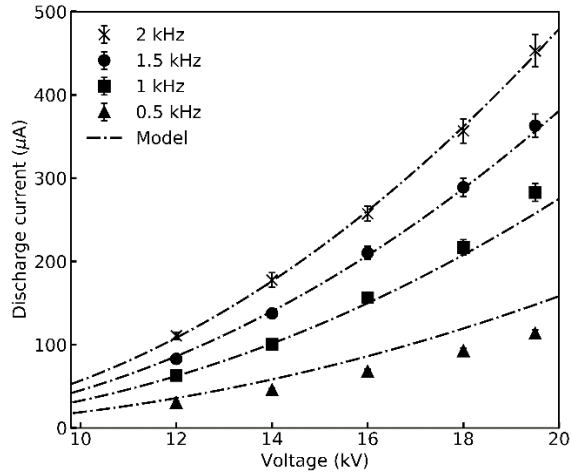

**Figure. 5.** Discharge current as a function of applied voltage and frequency for the experimental data.

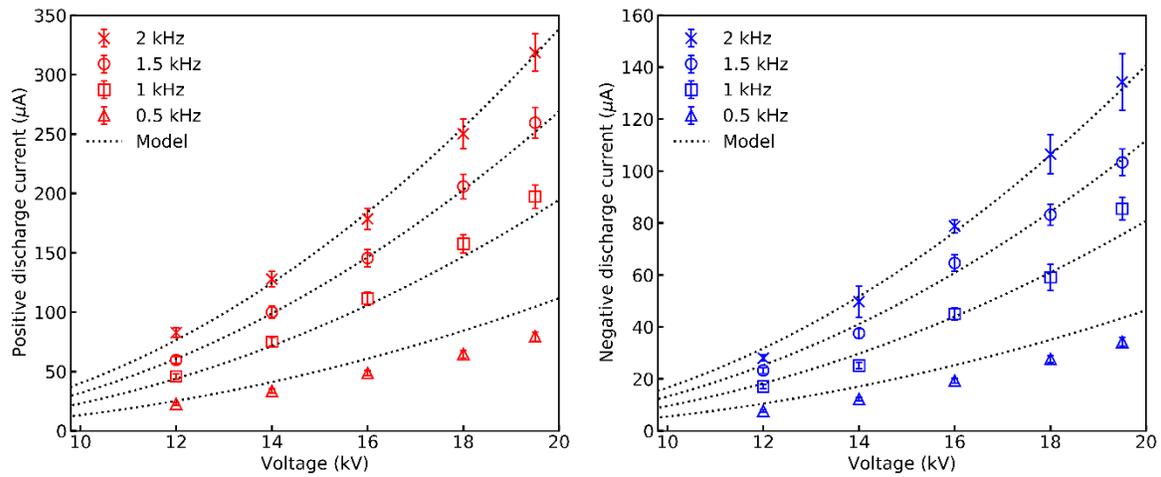

**Figure. 6.** Discharge current as a function of applied voltage and frequency for the experimental data. The current varies quadratically with applied voltage, and the power-law relationships for positive and negative discharge current predict the values accurately.

The relationships for the discharge currents are obtained by comparing the experimental results and using the expression in Eq. (5). The $I_{dis}$ versus $\varphi$ trends are similar to previously reported quadratic trends in the literature [17, 66, 67]. As expected, the discharge current increases when the frequency increases since the number of cycles increase with the increase in frequency. However, the discharge current cannot continuously increase with the increase in frequency because the build-up of charges on the dielectric surface tends to dampen the charge transport. The model agrees with the experimental results at all voltages and frequencies within ~10% error. The large asymmetry is due to both the number of current pulses in each cycle and the different mean charge transported by a single discharge[39]. The positive and negative discharge current models as shown in Eq. (14) and Eq. (15) are used to compare the empirical model and experimental data

$$I_{dis\_PD} = K_3 f_{AC}{}^{\alpha}(\varphi - \varphi_0)^2, \quad (14)$$

$$I_{dis\_ND} = K_4 f_{AC}{}^{\alpha}(\varphi - \varphi_0)^2, \quad (15)$$

where $I_{dis\_PD}$ and $I_{dis\_ND}$ are the discharge currents associated with positive and negative discharge, $K_3$ and $K_4$ are the empirical experimental constants. In our experiments, the ratio of the $K_3$ and $K_4$ is 2.5 and the positive discharge current is 2.5 times greater than the negative discharge current, as shown in Figure. 6.

We calculate the current density from the discharge current and plasma volume. In Figure. 7, the charge densities are plotted for 1 kHz and 2 kHz at varying voltages from 12 kV – 19.5 kV. The current density increases linearly with voltage in our experimental range. However, the current density is independent of frequency. The empirical relationships developed in the present study can be used to determine the charge density. In previous work, the simulation's charge density input is "tuned" to match the velocity profiles measured experimentally using an iterative approach [68]. The discharge current and discharge volume computed in this manuscript can be used as input parameters for a numerical model. The expressions given in Eq. (14) and Eq. (15) can be used to understand the momentum transfer process in positive and negative cycles.

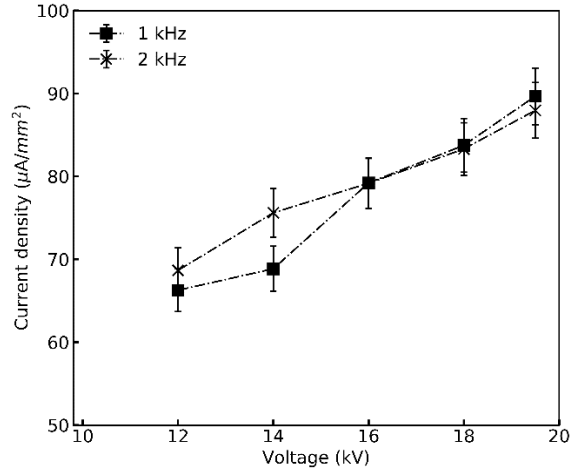

**Figure. 7.** Current density for total discharge current and the current density remains the same for the two applied frequencies.

### 3.3. Velocity characteristics

The velocity profiles of the DBD wall jet are used to determine the momentum injection. Figure. 8 shows the velocity profiles at different x locations for 14 kV and 18 kV at 1 kHz and 2 kHz. In all cases, the profile nearest to the active electrode shows the highest maximum velocity, $U_{max}$ and the greatest velocity graidient, i.e., shear stress. The flow is accelerated in the plasma discharge region [69], and as the wall jet propagates downstream, the velocity profile flattens due to viscous losses, momentum displacement [42], and specific to the EHD scenario, due to space charge effects. This jet-like expansion from the active electrode has been shown in similar pitot-tube experiments and the PIV experiments, and some have modeled DBD actuators using wall-jet similarity [26, 42, 43]. The plasma lengths are less than 10 mm in all cases, so we estimate that at 15mm, the EHD forcing is significantly reduced, and the velocity profiles are influenced mostly by viscosity and boundary layer momentum displacement. This is analogous to the analysis of EHD-driven flow by Guan et al. [25]

where the ratio of Coulombic force and an inertial term in the Navier-Stokes equation is presented as a non-dimensional parameter $X$. In the region of $X >1$, the flow is dominated by the Couloubic force, i.e., accelerating, and for $X \ll 1$, the EHD forcing can be neglected. For the cases presented in Figure. 8 (d) 18 kV at 2 kHz, the max velocity is 4.30 m s$^{-1}$ at y = 0.5 mm and x = 10 mm. The location and magnitude of maximum velocity agree with other experimental studies [1, 27, 70]. Some previous work report $U_{max} \sim 5.0$ m s$^{-1}$ at similar electrical inputs conditions, however, differences associated with electrode configuration and measurement methods can influence the $U_{max}$ measurements [27]. Also, PIV measurements show steep velocity gradients in the wall jet, thus, very slight difference in the probe position or finite dimention of the pitot tube (ID = 0.4 mm) may also cause lower $U_{max}$ values [26, 71]. Charging the of pitot tube could also adversely affect the electric field, thus decreasing the maximum velocity at the pitot tube tip. In the case of PIV measurements behavior of the highly charged particles in strong electrical field [72] may influence velocity measurements.

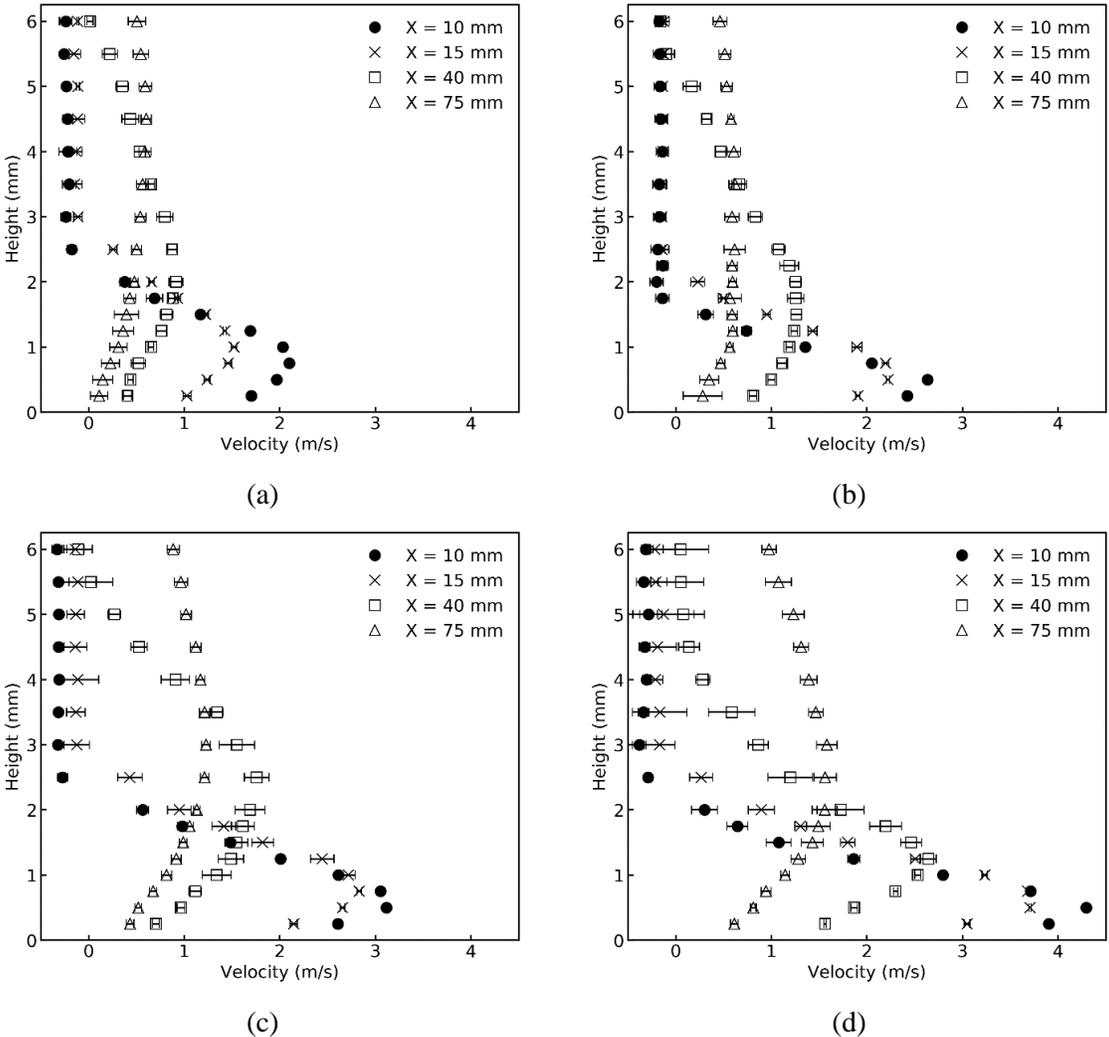

**Figure. 8.** Wall jet velocity profiles induced by DBD at different locations downstream from the high voltage electrode for different applied conditions (a) 14 kV / 1 kHz, (b) 14 kV / 2 kHz, (c) 18 kV / 1 kHz , and (d) 18 kV / 2 kHz.

An interesting and unreported phenomenon with pitot tubes is seen consistently in the velocity profiles. Negative velocities are observed above the wall jet region (y > 3.0 mm) at downstream locations of x = 10 mm and x = 15 mm, at downstream locations of x = 40 mm and x = 75 mm the x velocities are always positive, see Figure. 8. The magnitude of the negative velocity is larger at x = 10 mm than x = 15mm. One possible explanation is that the fluid entrainment by the DBD creates an adverse pressure gradient near the virtual origin that entrains the fluid from all directions, including from the downstream region. This trend was observed in PIV measurements by Debien [59]; however, it has not been reported in pitot tube measurements, such as Benard [5]. At this time, we do not have a satisfactory explanation of why the negative velocity was not observed in the other pitot tube experiments; perhaps, modeling of these flow cases can shed insight into the flow patterns.

The $U_{max}$ increases approximately linearly with voltage in this range of operating conditions. Other studies such as Forte [27] have shown a similar relation nearly linearly in the voltage range of 2 kV – 26 kV. Frequency also increases $U_{max}$, however, works such as Forte [27] and Jolibois [70] have shown that $U_{max}$ eventually reaches an asymptotic limit at increasing frequencies. Some reports suggest that the initial increase in $U_{max}$ is due to ion generation per cycle remaining constant as the frequency increases leading to increasing total ion generation, and momentum [5]. The plateau is attributed to a short enough AC voltage cycle that the generated ions on the dielectric surface between two discharge cycles do not have enough time to relax and transfer its momentum into an ionic wind [5]. Due to the uncertainty in y-position of the probe integration of the velocity profile over probe diameter, further experiments would be beneficial to elucidate the $U_{max}$ trends.

### 3.4. Energy transfer characteristics

Figure. 9 illustrates the energy conversion efficiency calculated as the ratio of mechanical power (or kinetic energy flux in the flow) to the electrical power as shown in Eq. (11) for 1 kHz and 2 kHz. The plasma actuator's electrical power consumption is calculated using Eq. (2). At the lowest electrical power consumption configuration of 14 kV and 1 kHz, the 110 mm spanwise electrode's power consumption was 3.34 W, which translates to approximately 0.304 W/cm. At the highest power consumption configuration of 18 kV and 2 kHz, the power consumption was 12.8 W, which translates to approximately 1.17 W/cm. These electrical power consumption levels support previous data of ~ 1 W/cm [5]. Similar to the momentum calculations, the kinetic energy flux is calculated using Eq. (10) at x = 15 mm. At both frequencies, the mechanical power increased with voltage.

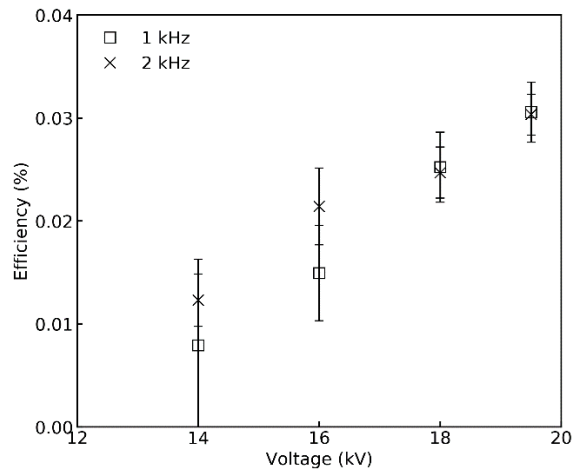

**Figure. 9.** Electromechanical efficiency determined at different voltages for two different frequencies reaches a maximum value of 0.03%.

From equation Eq. (11), the maximum electromechanical efficiency is ~ 0.03% at 19.5 kV and 1 kHz. The DBD's lowest efficiency of ~0.008% is the mildest conditions (14 kV and 1 kHz). In this voltage and frequency range, the efficiency continued to increase approximately linearly with voltage.

The calculated efficiency values agree with previous studies of traditional DBDs with thin dielectrics [70]. Other studies have shown that electromechanical efficiency reaches an asymptotic limit with higher electrical power input, and factors such as dielectric thickness and material properties affect the overall electromechanical efficiency [5]. For example, Laurentie et al. [60] studied the effect of electrode encapsulation and reached the efficiency values of 0.2%. Optimization of electrical input characteristics has shown promising results in increasing the electromechanical efficiency [73].

### 3.5. Relationship between discharge current and momentum

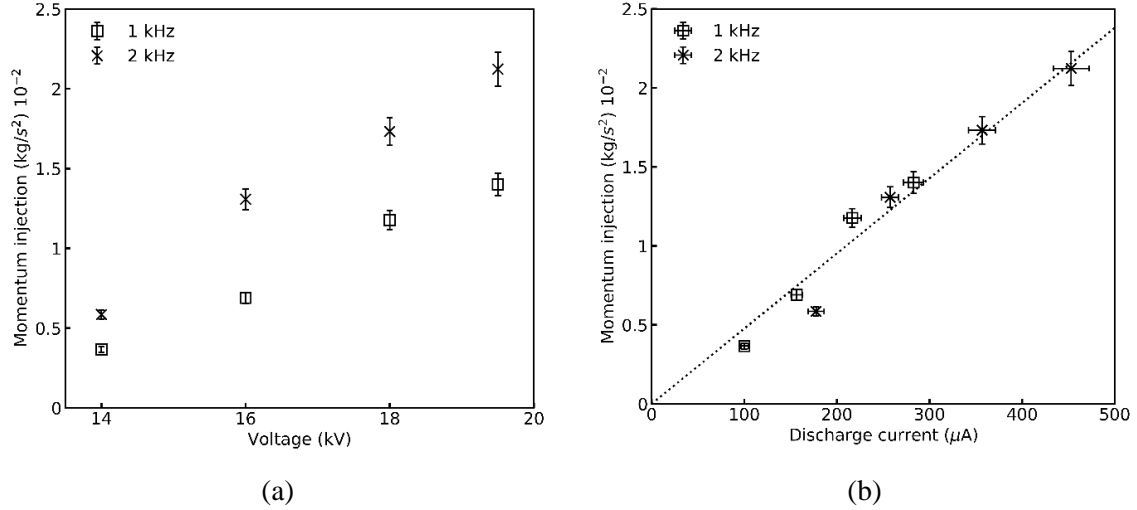

(a) (b)

**Figure. 10.** Momentum of the wall jet vs. (a) voltage and (b) discharge current. The wall-jet momentum collapses onto a single line for both frequencies and is directly proportional to the discharge current.

Analysis of wall jet momentum arguably is more important for the characterization of the wall jet. It is also less susceptible to errors associated with probe positioning and geometry as multiple data points are taken at each x location. Figure. 10(a) shows the DBD wall jet momentum calculated using equation (9) at x = 15 mm. The x = 10 mm location is not used; for 19.5 kV condition, the 10mm case had lower velocities than 15 mm. This is likely due to (i) probe interaction with plasma volume (plasma length ~ 7 mm), (ii) the flow was still accelerating (X >1). Excluding the 19.5 kV case, the momentum calculated at x = 10 mm is nearly identical to the momentum calculated at x = 15 mm. The momentum increases with the applied voltage and frequency for all studied conditions.

To understand the relationship between the fluid dynamic and plasma discharge, the induced wall jet momentum induced is plotted against the voltage and discharge current in Figure. 10. As expected, the increase in voltage and frequency leads to higher momentum. We do not have sufficient data to quantify the trend. However, the comparison of momentum and total discharge current combines both frequency and the voltage effects, and it reduces the data to a single line. In this analysis, only 8 data points are included over a relatively narrow range of frequencies. The least-square fit shows that the momentum increases linearly with the discharge current as shown in Figure. 10(a), and it can be approximated by Eq (16). From a mechanistic perspective, this is intuitive as the larger plasma volume, higher ion concentration, and stronger E-field lead to a greater number of ion-neutral molecule collisions accelerating the flow. The relationships linking frequency and voltage to discharge current are presented in Eq. (5). The obvious observation is that the planar DBD actuator can produce a higher momentum discharge current is maximized. The reduced-order model for DBD wall jet momentum is given by the following expression

$$M = \beta K f_{AC}{}^{\alpha}(\varphi - \varphi_0)^2. \tag{16}$$

The coefficients of this model are likely to depend on the electrode configuration and the properties of the dielectric media. It is also possible that the assumption of momentum being directly proportional to discharge current would break down at higher frequencies [64], and the exponents in the power-law relationship would change. At this time, we do not have sufficient data to extrapolate the model to these conditions. Table. 1 below summarizes the formulation of the reduced-order model for asymmetric DBD actuator developed from the experimental measurements for the range $\varphi = 12 – 19.5$ kV, $f_{AC} = 0.5 – 2$ kHz

**Table. 1** Summary of empirical expressions for different properties of DBD actuator

| Property | Expression | Coefficients |
|---|---|---|
| Plasma length ($L_P$) | $K_1 f_{AC}{}^{\alpha_1}(\varphi - \varphi_0)$ | $K_1 = 2.46 \times 10^{-2}$ $\alpha_1 = 0.38$ $\varphi_0 = 4.75$ kV |
| Plasma voume ($V_P$) | $K_2 f_{AC}{}^{\alpha}(\varphi - \varphi_0)^{\gamma}$ | $K_2 = 1.29 \times 10^{-4}$ $\alpha = 0.8$ $\gamma = 1.67$ |
| Total discharge current ($I_{dis}$) | $K f_{AC}{}^{\alpha}(\varphi - \varphi_0)^2$ | $K = 4.71 \times 10^{-3}$ |
| Positive discharge current ($I_{dis\_PD}$) | $K_3 f_{AC}{}^{\alpha}(\varphi - \varphi_0)^2$ | $K_3 = 3.33 \times 10^{-3}$ |
| Negative discharge current ($I_{dis\_ND}$) | $K_4 f_{AC}{}^{\alpha}(\varphi - \varphi_0)^2$ | $K_4 = 1.38 \times 10^{-3}$ |
| Momentum ($M$) | $\beta K f_{AC}{}^{\alpha}(\varphi - \varphi_0)^2$ | $\beta = 4.76 \times 10^{-3}$ |

## 4. CONCLUSION

We developed a reduced-order model for discharge current, current density, and momentum injection utilizing the data from DBD actuator, i.e., plasma volume, electrical discharge current, and resulting velocity profiles over a range of voltage (12 kV – 19.5 kV) and frequency (0.5 kHz – 2 kHz). The plasma length increases linearly with voltage, matching other previous studies, and the plasma volume is found to vary quadratically with voltage. The plasma volume continues to grow with a frequency up to 2 kHz, whereas the plasma length approaches a limit at 2 kHz at these operating conditions. The increase in volume is due to continued growth in plasma height, suggesting that plasma volume can be a better input in CFD modeling.

The current associated with microdischarges was measured using a Rogowski coil with high temporal resolution. The charge transported in each microdischarge and the corresponding discharge current was calculated for both PD and ND semi-cycles. The discharge current analysis yielded a power law for the positive and negative discharge current associated with microdischarges in the form $I_{dis} = K f^{\alpha}(\varphi - \varphi_0)^2$. Comparing the expressions, there is an asymmetry in the discharge currents between positive and negative cycles. The current density was calculated using discharge current and plasma volume. The discharge current density increases with voltage, and it is independent of frequency.

The momentum and mechanical power of the DBD actuator were determined using vertical velocity profiles at different downstream positions over a range of operating conditions. DBD wall jet momentum increases with voltage and frequency, and it is directly proportional to the discharge current. The electromechanical efficiency increases with voltage; maximum efficiency of ~0.03% agrees with previous data from thin dielectric DBD actuators. The discharge current expressions and baseline velocity were used to develop a reduced-order model of DBD momentum injection. And the analysis of the plasma volume can be used in the multiphysics modeling of the DBD. Future research should extend the study to higher frequencies to develop a more robust relationship between the discharge current and plasma volume to determine the body force acting on the fluid. Another topic is determining the interaction between the free flow and the DBD momentum injection into the flow boundary layer.

## 5. ACKNOWLEDGMENTS

This work was supported through an academic-industry partnership between Aerojet Rocketdyne and the University of Washington funded by the Joint Center for Aerospace Technology Innovation (JCATI) and is also based upon work supported in part by the Office of the Director of National Intelligence (ODNI), Intelligence Advanced Research Projects Activity (IARPA), via ODNI Contract 2017-17073100004. The views and conclusions contained herein are those of the author and should not be interpreted as necessarily representing the official policies or endorsements, either expressed or implied, of ODNI, IARPA, or the U.S. Government.

## NOMENCLATURE

| Symbol | Description |
|---|---|
| $V_P$ | Projected volume of the plasma zone |
| $C$ | Pitot tube correction factor |
| $E$ | Electric field |
| $f_{AC}$ | Frequency of the applied voltage |
| $\vec{f}_{EHD}$ | Electro-hydrodynamic force term |
| $i(t)$ | Current |
| $I_{dis}$ | Discharge current |
| $K_i$ | Experimental constants |
| $L$ | Spanwise Length |
| $L_P$ | Length of the plasma region |
| $M$ | Momentum of the induced jet |
| $P$ | Pressure reading from the pitot tube |
| $W$ | Discharge energy consumption |
| $W_{mech}$ | Mechanical power |
| $W_{elec}$ | Electrical power |
| $U(y)$ | Velocity at y height |
| $U_{max}$ | Maximum velocity of the wall jet |
| $v$ | Time-averaged velocity |
| $\alpha_i$ | AC frequency factors |
| $\beta$ | Proportionality constant between discharge current and momentum |
| $\gamma$ | AC high voltage factor for plasma volume |
| $\varphi(t)$ | Peak to peak AC high voltage |
| $\varphi_0$ | Initiation voltage peak to peak |
| $t^*$ | Normalized time value |
| $\rho$ | Density |
| $Q$ | Mass flow rate |
| $\eta$ | Efficiency |